\documentclass[review]{elsarticle}

\usepackage{lineno,hyperref}
\modulolinenumbers[5]

\journal{Elsevier}

\usepackage{graphicx}
\usepackage{epstopdf}
\usepackage{dcolumn}
\usepackage{color}
\usepackage{multirow}
\usepackage{rotating}
\usepackage{array}
\usepackage{bm}

\newcommand{\be}{\begin{equation}}
\newcommand{\ee}{\end{equation}}
\newcommand{\ba}{\begin{eqnarray}}
\newcommand{\ea}{\end{eqnarray}}
\newcommand{\bpm}{\begin{pmatrix}}
\newcommand{\epm}{\end{pmatrix}}
\newcommand{\bs }{\boldsymbol }
\newcommand{\ver}{\bs{r}}

\begin{document}

\begin{frontmatter}
\title{Testing the formation of Majorana states using the Majorana Polarization}

\author{Cristina Bena}
\address{Institut de Physique Th\'eorique, Universit\'e Paris Saclay, CEA, CNRS,
Orme des Merisiers, F-91191 Gif-sur-Yvette, France}

\begin{abstract}
We study the formation of Majorana states in superconductors using the Majorana polarization, which can locally evaluate the Majorana character of a given state. We introduce the definition of the Majorana polarization vector and the corresponding criterion to identify a Majorana state, and we apply it to some simple cases such as a one-dimensional wire with spin-orbit coupling, subject to a Zeeman magnetic field, and proximitized by a superconductor, as well as to an NS junction made with such a wire. We also apply this criterion to two-dimensional finite-size strips and squares subject to the same physical conditions. Our analysis demonstrates the necessity of using the Majorana polarization local order parameter to characterize the Majorana states, particularly in finite-size systems.
\end{abstract}

\end{frontmatter}


\section{Introduction}
Majorana fermions have been attracting a lot of interest recently ; these atypical fermionic particles have been predicted long time ago by E. Majorana as real solutions of the Dirac equation \cite{Majorana2008}. Regular fermions that have each their corresponding antiparticle, for example in condensed matter physics the antiparticle of an electronic excitation is a hole. However the Majorana particles are their own antiparticles, which makes for a very special and intriguing situation. In nature there are very few instances of finding Majorana particles. Thus, neutrinos are believed to be Majorana fermions, and while their properties are consistent with such an assumption, a definite experimental proof is still lacking. In condensed-matter systems there is no a priori motivation to look for Majoranas, since such systems are composed of regular fermions, and for a long time no search for Majoranas was conducted in such systems. However, over the past years, it has been proposed that even condensed-matter systems may support Majorana-like excitations, in the form of collective modes, in a similar manner in which electronic systems may support fractionally charged excitation. Thus, the first reference to such a realization of Majoranas was given by Kitaev in 2001\cite{Kitaev2001}, followed subsequently by a large number of works proposing different condensed-matter platforms supporting Majoranas (see \cite{Alicea2012} for a review).

A simple manner to think about the possibility of the realization of Majorana states in a condensed matter system is to imagine a setup in which the elementary excitations are equal combinations of electrons and holes. Thus the reality condition, implying that a given excitation is identical to its hermitian conjugate, is automatically satisfied. Such excitations made of combinations of electrons and holes appear naturally in superconductors, and thus we expect that the superconductors would be the ideal ground to start looking for Majoranas. However,  in regular s-wave superconductors the natural excitations (Bogoliubov qausiparticles) are made up of electrons and holes with opposite spins, that are not self-adjoint due to their spin structure. To have a true self-adjoint state one would require a p-wave superconductor, but such superconductors are rare and their p-wave character not fully agreed upon (e.g. $Sr_2 Ru O_4$ \cite{Mackenzie2003}). One would therefore need a better method to realize such excitations.

Thus, among the most studied Majorana proposals, both theoretically and experimentally, are one-dimensional (1D) systems with a strong spin-orbit coupling such as  InAs and InSb  wires \cite{NadjPerge2013,nilsson2009}, subject to Zeeman magnetic field and proximitized by a superconductor (SC); it has been shown that such systems can exhibit Majorana bound states (MBS) at their extremities \cite{Lutchyn2010,Oreg2010}. Concerning the experimental confirmation, though there have already been several promising experiments \cite{Mourik2012,Deng2012,Das2012,Lee2014,Nadj-Perge2014} their interpretation is controversial and no clear-cut detection of Majoranas has so far been agreed upon.

The goal of the present work is to introduce a new tool to study Majorana fermions in condensed matter systems. We show here that a fundamental ingredient in studying these excitations is lacking, and we present this missing piece and prove its importance. We thus introduce the generalized Majorana polarization (MP) \cite{Sedlmayr2015b,Sticlet2012}, which is a universal measure of the spatial dependence of the Majorana character of a given state, i.e. of the same-spin particle-hole overlap. This quantity can be thought of as the analog of the local density of states (LDOS) for regular electrons, except that while the LDOS measures the number of available electronic states, the MP measures the number of available Majorana states at a given energy and position. Moreover, as we will show, a real quantity does not suffice to capture the Majorana character, and one needs to introduce a complex quantity which can be represented as a  two-component vector in the complex plane.

Having access to such a local measure can allow one to understand the evolution of the Majorana states through a phase transition, their dependence on specific particularities of the system such as size, disorder, inhomogeneities, etc., as well as how one can manipulate them. 

We show that the calculation of the MP allows one to write down a topological criterion and assign a global topological character to a given state. When the criterion of zero energy for a given state cannot be strictly applied (e.g infinitesimally small but non-zero energies), having access to a local order parameter is a sufficient and versatile criterion for such a distinction. This allows for an accurate determination of the topological phase diagram from numerical calculations.

In what follows we write down the generalized MP definition and apply it to a few examples, such as a one-dimensional wire with Rashba, Zeeman and SC proximity, as well as to NS junctions made with such wires. Also, we apply it to two-dimensional finite-size strips and squares with open boundary conditions. 

The paper is organized as follows: In section 2 we present the  Majorana polarization, in section 3 we introduce the most general model Hamiltonians for the systems studied, in section 4 we apply the MP criterion to the simple case of a one-dimensional wire, in section 5 we consider an NS junction, while in section 6 we study two-dimensional finite-size systems. We conclude in section 7.

\section{Majorana Fermions and the Majorana Polarization}

Since Majorana fermions are their own anti-particles $\gamma^\dagger=\gamma$, where $\gamma^\dagger$ is the creation operator for a Majorana fermion. In condensed-matter systems in which the elementary excitations are electrons ($c^\dagger$) and holes ($c$), Majorana states can be achieved as collective excitations, in particular as superpositions of electronic and hole states ($\gamma=(c\pm c^\dagger)/\sqrt{2}$), superpositions that can satisfy the self-adjoint-conditions. Such superpositions can occur naturally in superconductors, with equal superpositions of electrons and holes being predicted at zero-energy. However, regular superconductors would pair electrons and holes of opposite spins ($v c_\uparrow + u c^\dagger_\downarrow$), while Majorana states require superpositions of electrons and holes with the same spin  ($\gamma_\alpha=c_\alpha\pm c^\dagger_\alpha)/\sqrt{2}$). It has been proposed that they can arise naturally in one-dimensional wires with strong spin-orbit coupling in the presence of a Zeeman field and in the proximity of a regular superconductor.

To quantify such superpositions we introduce the Majorana polarization \cite{Sedlmayr2015b,Sticlet2012}. This is an analogous quantity to the LDOS that measures the number of electronic states available as a function of energy and position. The MP measures the number of Majorana states available as a function of the energy and position. Moreover, for a single state, the LDOS can provide information about the distribution of the electronic density as a function of position, and consequently indicate where the electronic state is more likely to be localized. In a similar manner the MP of a given state can tell where the Majorana state is more likely to be localized. However, while the LDOS is a scalar quantity, the MP is a vector quantity. This is because to describe most generally the electron-hole overlap of a given state a scalar quantity does not suffice. The most general Majorana state can be written as $\gamma=(c + e^{i \theta}c^\dagger)/2$, where $\theta$ is an arbitrary phase. Such combination can be characterized by a scalar quantity capturing the amount of electron-hole overlap, but also by a phase, necessary to describe the relative phase between the electron and a hole component. This phase makes the MP a vector that spans the two-dimensional complex plane, with a magnitude between zero and one, zero for purely electron or hole states, and one for purely Majorana states. We can think about the MP vector as a pseudo-spin polarized LDOS, with the pseudo-spin direction corresponding to the relative phase between the electron and the hole component.

\begin{figure}
\center
\includegraphics*[width=0.6\columnwidth]{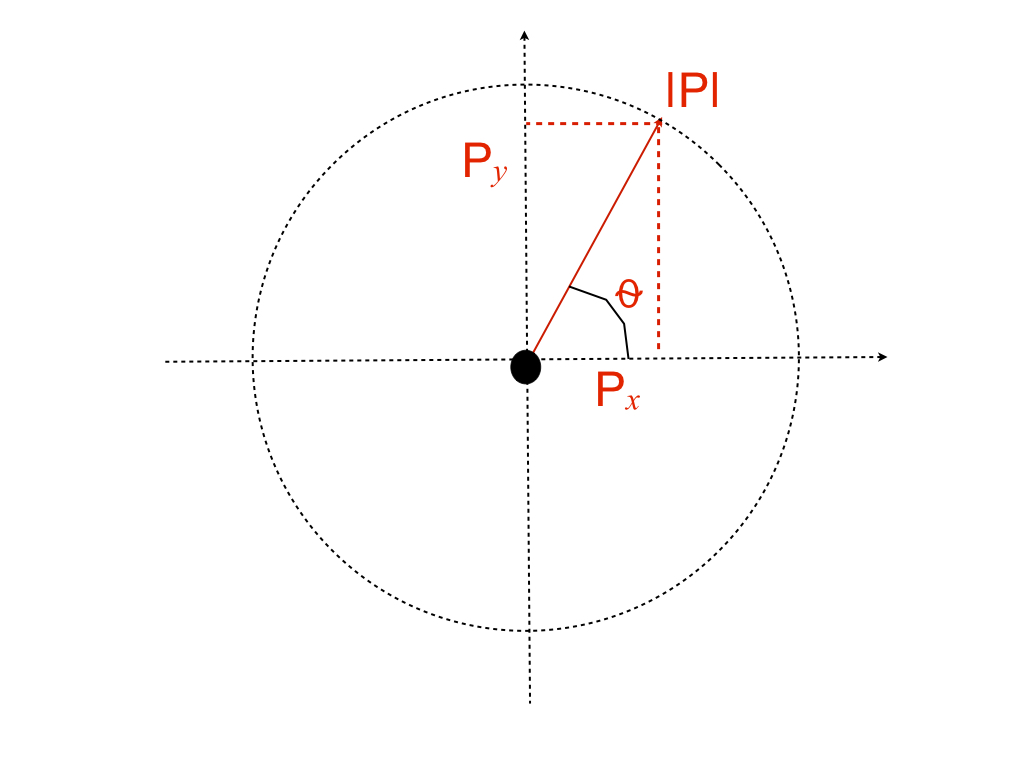}
\caption{(Color online) The Majorana polarization vector $P$ in the complex plane.}
\label{fig1}
\end{figure}

To take a more specific example we consider first a spinless model for which the most general solution for a given state is a combination of electrons and holes $\Psi=u c^\dagger + v c$. We can see here that such a state is purely electronic or purely hole if either $u$ or $v$ is zero, and it is a Majorana if $|u|=|v|=1/2$. We introduce the Majorana polarization vector that characterizes the electron-hole overlap as 
\begin{equation}
P= 2 u v,
\end{equation}
or equivalently, in vector notation $P=(P_x,P_y)$ with $P_x=2 {\rm Re} (u v)$ and $P_y=2 {\rm Im} (u v)$.
We note that $P$ is indeed zero for either $u=0$ or $v=0$, while for a Majorana state, for which $|u|=|v|=1/2$, or equivalently $\Psi_M=e^{i\phi}(c^\dagger + e^{i \theta} c)/2$, then $P=e^{2 i \phi} e^{i \theta}$. This has indeed the absolute value of $1$. The resulting phase has two components, $e^{2 i \phi}$ comes from the overall arbitrary phase factor of the wave function and it is generally irrelevant, and can be gauged away. The second phase factor $e^{i \theta}$ has however a physical significance since it captures the information about the relative phase between the electron and hole component. As we will see in what follows it does not have to be spatially uniform and its spatial structure is crucial in the study of a given Majorana state. The simplest example underlying the importance of this phase factor is that of a one-dimensional wire in which the right-end Majorana has $\theta=\pi$, while the left-end one has $\theta=-\pi$ (in the literature the corresponding Majoranas are usually denoted $\gamma_{1/2}=(c^\dagger \pm c)/\sqrt{2} $). We should note that for any given closed system the distribution of $\theta$ needs to satisfy the condition that the integral   of the MP vector over the entire system \footnote{Here, and in what follows, by the 'integral' of the MP vector quantity over a given region we denote the absolute value of the sum of the MP vectors for all the lattice sites inside the corresponding region.}is equal to zero (there cannot exist free-standing Majorana monopoles in nature). The phase $\theta$ can be usually related to the SC phase, but it can be influenced also by many factors and the details of the model and of the setup. While $|P|=1$ corresponds to a Majorana state, $|P|<1$ is not a Majorana though a state with a large value of $|P|$ may also have some interesting properties. A state with $|P|=0$ is a pure electron or hole state. A graphic description of the MP is presented in Fig.~\ref{fig1}.

For a spinful model for which a given wavefunction has the most general form $\Psi=u_\uparrow c_\uparrow^\dagger + v_\uparrow c_\uparrow+u_\downarrow c_\downarrow^\dagger + v_\downarrow c_\downarrow$ the MP can simply be generalized to

\begin{equation}
P= 2 u_\uparrow v_\uparrow+2 u_\downarrow v_\downarrow,
\end{equation}
or equivalently $P=(P_x,P_y)$, with $P_x=2 {\rm Re} (u_\uparrow v_\uparrow+u_\downarrow v_\downarrow)$ and $P_y=2 {\rm Im} (u_\uparrow v_\uparrow+u_\downarrow v_\downarrow)$.

\begin{figure}
\center
\vspace{-.1in}
\includegraphics*[width=0.5\columnwidth]{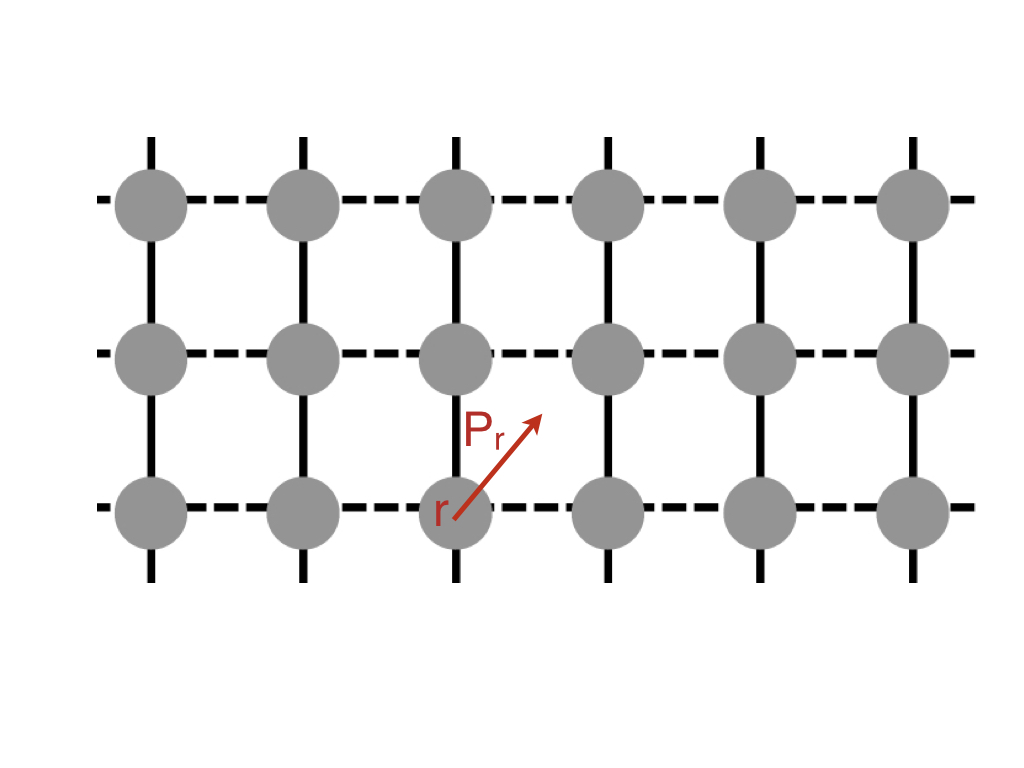}
\vspace{-.5in}
\caption{(Color online) The Majorana polarization vector represented on a site of a square lattice.}
\label{fig2}
\end{figure}

To understand how the definition of the MP works for an extended system we consider a two-dimensional lattice for which the MP needs to be defined on each site $\ver = (x,y)$ (see Fig.\ref{fig2}), for each eigenstate $\Psi_j(\ver)=u^j_{\ver \uparrow} c_{j \uparrow}^\dagger + v^j_{\ver \uparrow} c_{j\uparrow}+u^j_{\ver \downarrow} c_{j \downarrow}^\dagger + v^j_{\ver \downarrow} c_{j \downarrow}$,  and can thus be written as:
\begin{eqnarray}
P^j({\ver}) \equiv \big\{P^j_x({\ver}), P^j_y({\ver})\big\} 
\equiv \Big\{2 {\rm Re} \left[u^j_{\ver \uparrow} v^j_{\ver \uparrow} + u^j_{\ver \downarrow} v^j_{\ver \downarrow} \right], 2 {\rm Im} \left[u^j_{\ver \uparrow} v^j_{\ver \uparrow} + u^j_{\ver \downarrow} v^j_{\ver \downarrow} \right] \Big\} 
\label{MPvector}
\end{eqnarray}

Additionally, in a region $\mathcal{R}$ where such a Majorana state is localized, its wavefunction must satisfy
\begin{equation}\label{mp}
C=\frac{\left|\sum_{j \in \mathcal{R}}P^j(\ver)\right|}{\sum_{j\in \mathcal{R}}\langle\Psi_j(\ver)|\Psi_j(\ver)\rangle}=1\,,
\end{equation}
where $\langle\Psi_j(\ver)|\Psi_j(\ver)\rangle$ is the electronic density at position $\ver$ for the eigenstate $j$. For a system with two Majoranas localized each on a different edge of the system, $\mathcal{R}$ can simply be taken to be half the system.

To obtain the topological phase diagram we first find the lowest energy states of the given system. If these states have energies close to zero they may be MBS. We divide our system into two halves (along the shorter length), and we compute the integral of the MP vector in each of these halves defined by $\ver \in \mathcal{R}$ for each 'zero'-energy state. The states that have $C=1$ are MBS, and those with $C=0$  are regular electron or hole states. Note that we may have only a pair of state with $C=1$, or multiple degenerate zero-energy MBS states with $C=1$. 

\begin{figure}
\center
\vspace{-.5in}
\includegraphics*[width=1\columnwidth]{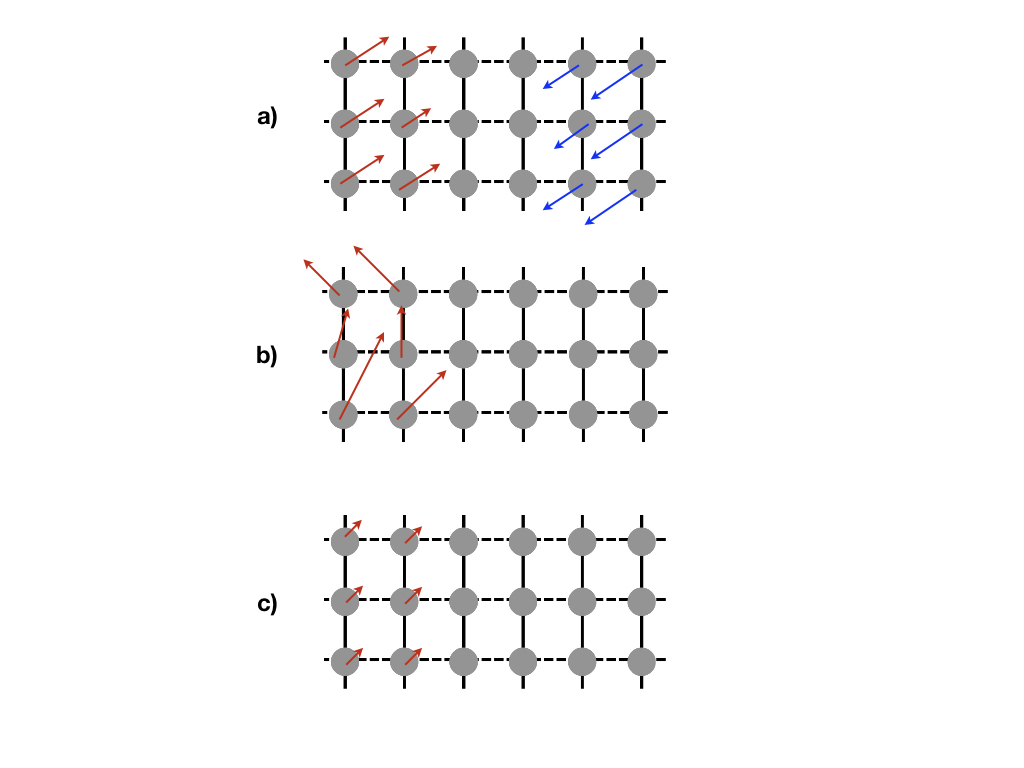}
\vspace{-.5in}
\caption{(Color online) A few typical MP configurations: a) Majorana state, MP large and aligned, b) quasi-Majorana state, MP large but non aligned, c) trivial state, MP=0.}
\label{fig3}
\end{figure}

A few examples of typical MP configurations are given in Fig.~\ref{fig3}. 
In Fig.~\ref{fig3}a) a typical Majorana setup is depicted, in which the two necessary conditions are satisfied: 1. over half of the system the MP vectors are aligned, and 2. locally the MP value is equal to the electronic density (local perfect Majorana character). These conditions need to both be satisfied in order that the sum of the MP vector over half the wire to be equal to the sum of the electronic densities. The total integral of the MP vector over the entire system is zero, as expected. In Fig.~\ref{fig3}c) we depict the trivial situation in which locally the MP vectors are very small (local non-Majorana character), this situation corresponds to a trivial phase. In Fig.~\ref{fig3}b) we depict an interesting situation that can arise in which locally one may have a perfect Majorana character (equal MP and electronic density), but the MP vectors are not aligned, thus yielding a sum smaller than the total electronic density. For this situation $C<1$ and the topological criterion is not satisfied, however the state, even if non-topological has interesting properties that we will discuss in the next sections, we call such a state a quasi-Majorana state.

In what follows we apply this definition and the MP criterion to test the formation of Majorana states in a few particular situations of one-dimensional and two-dimensional lattices.

\section{Model}

We consider a tight-binding Hamiltonian, with a Rashba spin-orbit  coupling $\alpha$, subject to a magnetic field $\vec{B}(\vec{r})$ and brought in the proximity of a superconducting substrate which is assumed to induce an on-site SC pairing.  We write down the corresponding Bogoliubov-de-Gennes Hamiltonian in the Nambu basis, $\Psi_{j}=(\psi_{j,\uparrow},\psi_{j,\downarrow},\psi^{\dag}_{j,\downarrow},-\psi^{\dag}_{j,\uparrow})^T$, where $j$ denotes the corresponding site. The full Hamiltonian is
\begin{equation}\label{hamiltonian}
 H=H_0+H_{\rm B}+H_{\rm R}\,.
\end{equation}
The first term of the Hamiltonian is given by
\begin{equation}
H_0=\sum_{j}\Psi^\dagger_j\left[-\mu{\bm\tau}^z-\Delta{\bm\tau}^x\right]\tilde\Psi_j
-t\sum_{\langle i,j\rangle }\Psi^\dagger_i{\bm \tau}^z\Psi_{j}\,,
\end{equation}
where $\mu$ is the chemical potential, $t$ the hopping strength, $\sum_{\langle i,j\rangle }$ denotes summing over nearest neighbors, and $\Delta$ the induced superconducting pairing. We set $t=\hbar=1$ throughout. We also take the distance between nearest neighbours $a=1$.

The second term in Eq.~\ref{hamiltonian} describes the effect of the Zeeman magnetic field of strength $B$ and is given by
\begin{equation}\label{hmag}
H_{\rm B}=\sum_j \Psi^\dagger_j\vec{B}_j\cdot\vec{\sigma} \Psi_j\,.
\end{equation}

While this is the most general form describing a magnetic field pointing in an arbitrary direction and that can vary spatially, in most of what follows we will consider a uniform field pointing in the $z$ direction.

The term corresponding to the effective Rashba spin-orbit interaction of strength $\alpha$ can be written as:
\begin{equation}\label{rashba}
H_{\rm R}={\rm Im}[\alpha\sum_{\langle i,j\rangle }\Psi^\dagger_i\left(\vec{\delta}_{ij}\times{\vec{\bm \sigma}}\right)\cdot\hat{z} {\bm \tau}^z\Psi_{j}]\,.
\end{equation}

\section{One-dimensional wire}

\begin{figure}
\center
\vspace{-1in}
\includegraphics*[width=.6\columnwidth]{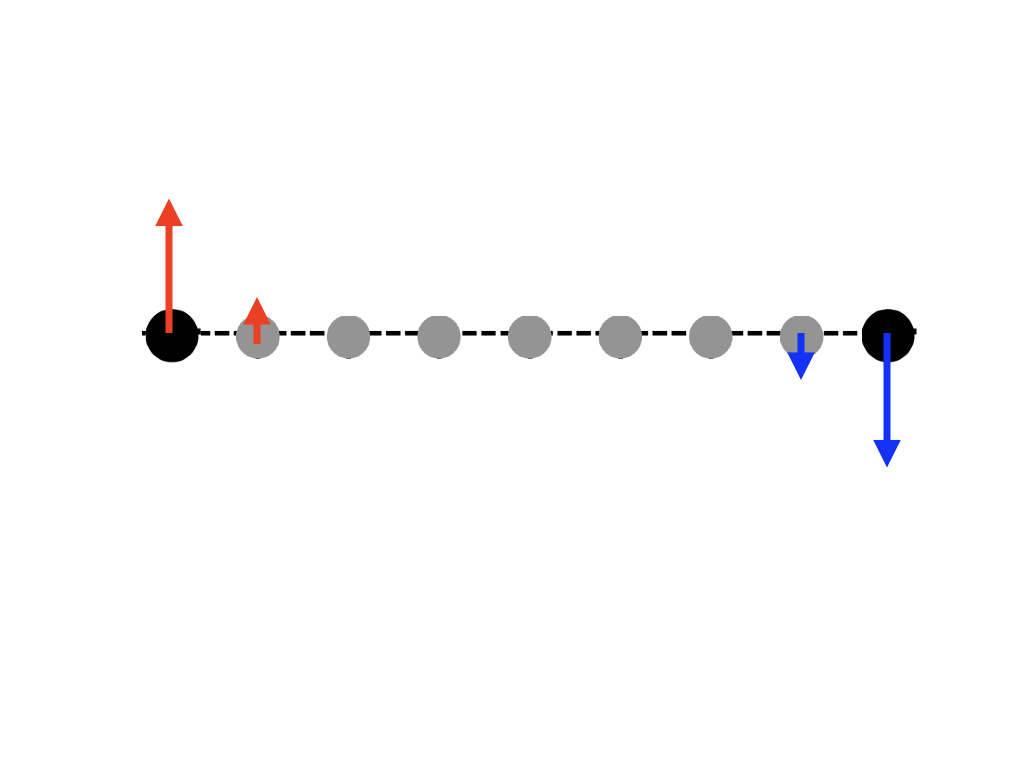}
\vspace{-1in}
\caption{(Color online) One-dimensional wire: Majorana states form with MP vectors pointing in opposite direction at the two ends.}
\label{fig4}
\end{figure}

\begin{figure}[h]
\centering
\includegraphics[width=0.65\textwidth]{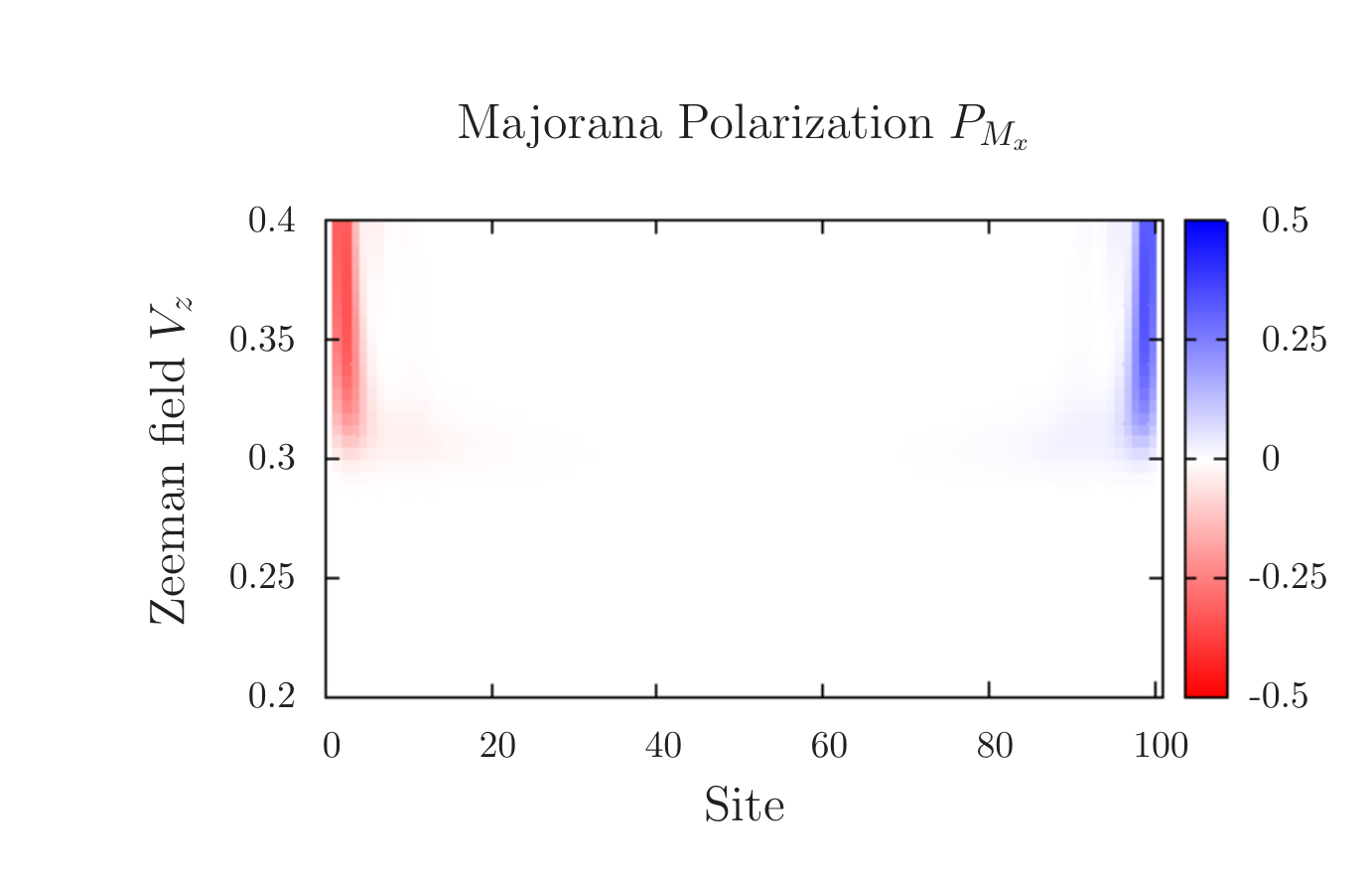}\\
\includegraphics[width=0.58\textwidth]{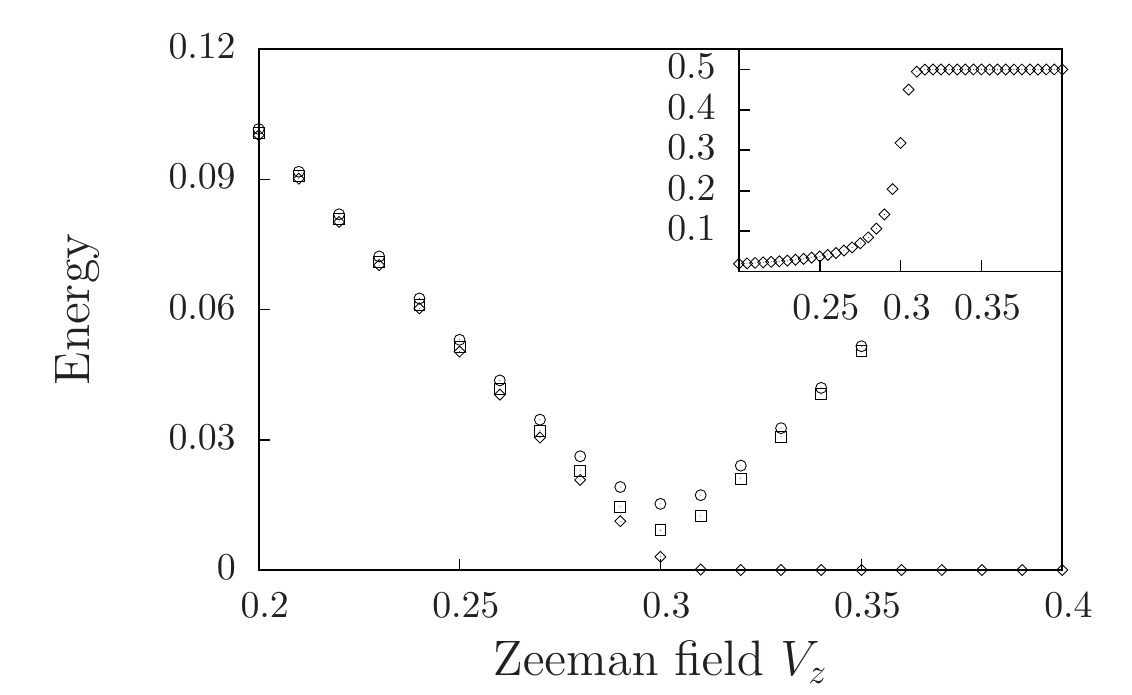}
\caption{\small Upper panel: Majorana polarization of the lowest-energy state as a function of position and $V_z$. Lower panel: lowest-energy eigenvalues and the half-wire integral of the Majorana polarization ($C/2$, inset) as a function of $V_z$.  Note that at $V_z=\Delta$ the MP goes to zero, and correspondingly the lowest-energy state acquires a non-zero value. Also $C=1$ for $V_z>\Delta$, and it decreases to zero sharply across the phase transition at $V_z=\Delta$. Parameters: $\Delta=0.3$, $\mu=0$, and $\alpha=0.2$}
\label{fig:majv}
\end{figure}

We first consider a one-dimensional wire (see Fig.~\ref{fig4}) and a uniform magnetic field pointing in the $z$ direction, $\vec{B}(\vec{r})=V_z \hat{z}$, and we study the formation of the Majorana end states. We perform an exact diagonalization of the tight-binding Hamiltonian described in the previous section and we obtain the eigenvalues and eigenfuctions of this system, as well as the MP of the lowest energy state. In Fig.~\ref{fig4} we depict schematically the configuration of the MP that we obtain, while in Fig.~\ref{fig:majv} we plot the MP of the lowest energy state as a function of the applied Zeeman field. We indeed recover Majorana bound states at the ends of the wire, with MP vectors pointing in the opposite direction\cite{Sticlet2012}. We note that the MP remains finite for $V_z>\Delta$ and goes to zero over the phase transition at $V_z=\Delta$ as expected.

Moreover, the lowest-energy states of the system stick to zero energy for $V_z>\Delta$ and acquire a non-zero value for $V_z<\Delta$ marking the phase transition between the topological and non-topological state. Also, as it can be seen from the inset of the lower panel Fig.~\ref{fig:majv}, the renormalized integral of the MP over half the wire ($C$) is equal to $1$ in the topological phase, and decreases abruptly to zero in the non-topological phase.  This allows us to confirm our assumption that the value of $C$ can be used as a good order parameter to describe a topological phase transition, and that the MP is a good local order parameter to describe the formation of Majorana states. 

\section{NS junctions}

\begin{figure}
\center
\vspace{-.5in}
\includegraphics*[width=.5\columnwidth]{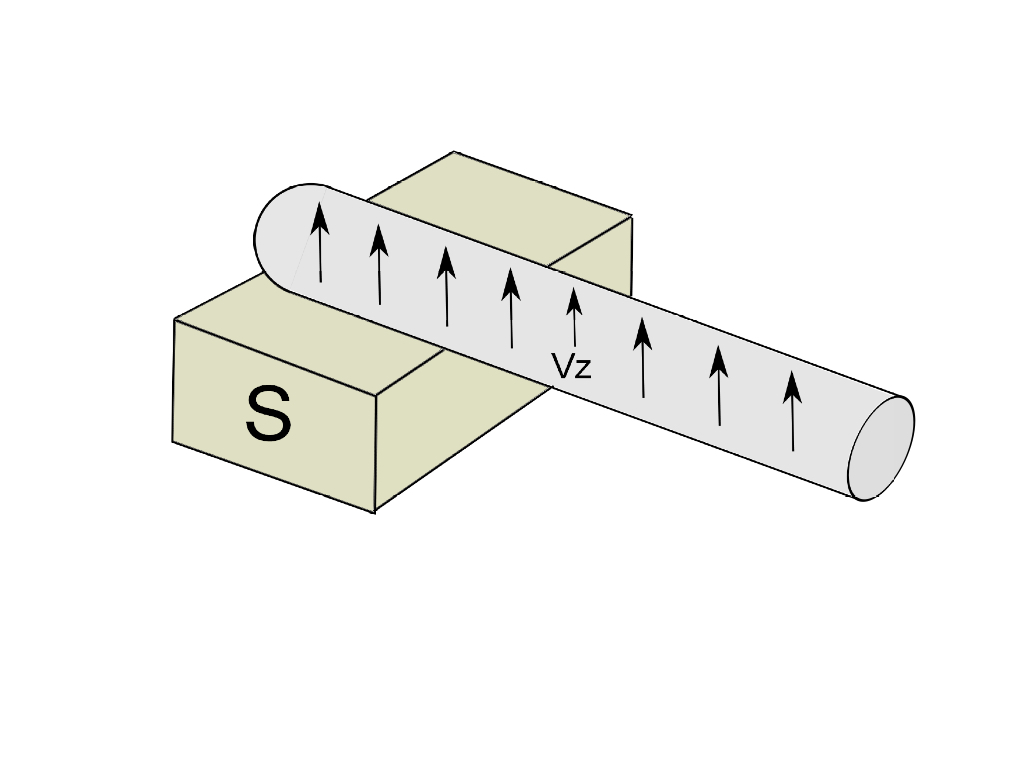}
\vspace{-.5in}
\caption{(Color online) Schematics of an NS junction made with a one-dimensional wire proximitized by a SC over half its length.}
\label{fig5}
\end{figure}

We focus on using a one-dimensional wire such as the one described in the previous section to make an SN  long junction. Thus, as described in Fig.~\ref{fig5}, the superconductivity is induced by proximity only in the section of the wire in contact with the superconductor. We model this by considering $\Delta_{x<20}=0.4t$ in the SC-proximitized region and $\Delta_{x\geq20}=0$ in the normal region.

\begin{figure}
\center
\vspace{-.5in}
\includegraphics*[width=0.7\columnwidth]{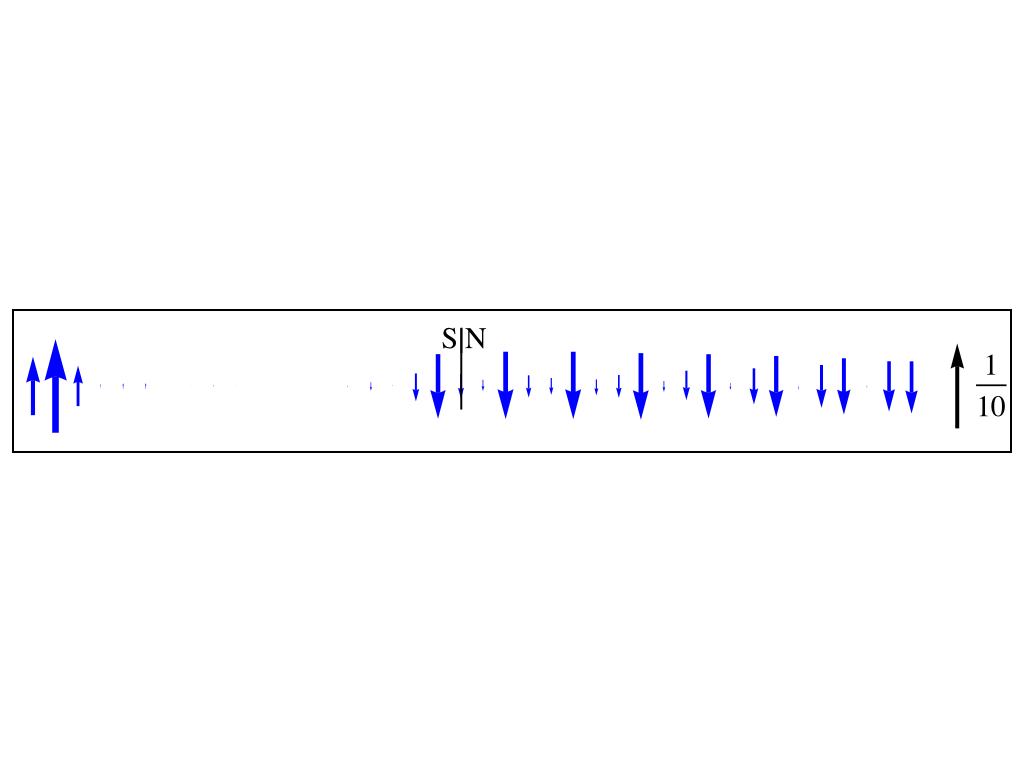}
\vspace{-1in}
\caption{(Color online) Local Majorana polarization of the lowest-energy state as a function of position for an NS junction with $t=1$, $\mu=0$, $\alpha=0.2t$, $V_z=0.5t$, $N=40$, and $\Delta_{x<20}=0.4t$ in the S region and $\Delta_{x\geq20}=0$ in the N region. The MBS contribution at the superconductor edge near $x\approx1$, has been scaled down by $1/4$ relative to the extended Majorana in the N region so that they can be shown in the same figure. }
\label{SN}
\end{figure}

By diagonalizing the Hamiltonian of this system we find that its spectrum also exhibits two zero-energy modes. In Fig.~\ref{SN} we focus on the example $t=1$, $\mu=0$, $\alpha=0.2t$, $V_z=0.5t$, $N=40$. We see that the zero-energy states correspond to a localized Majorana in the SC and an extended uniform Majorana in the normal state \cite{Chevallier2012}. We have checked that the integral of the renormalized MP for this state is exactly $1$, cancelling the MP of the edge mode in the SC (the total MP has to be conserved and equal to zero).

We note that the use of the MP local order parameter is crucial to understand the physics of this system. Without it one cannot say for sure if the zero-energy states are Majoranas, nor that the Majorana character is uniform and that the Majorana state extends thus over the entire normal region. Such observation about the localization of a Majorana state is very important for the detection and the manipulation of Majorana states in NS junctions.

\section{Finite-size strips and squares}

\begin{figure}
\center
\vspace{-.5in}
\includegraphics*[width=0.7\columnwidth]{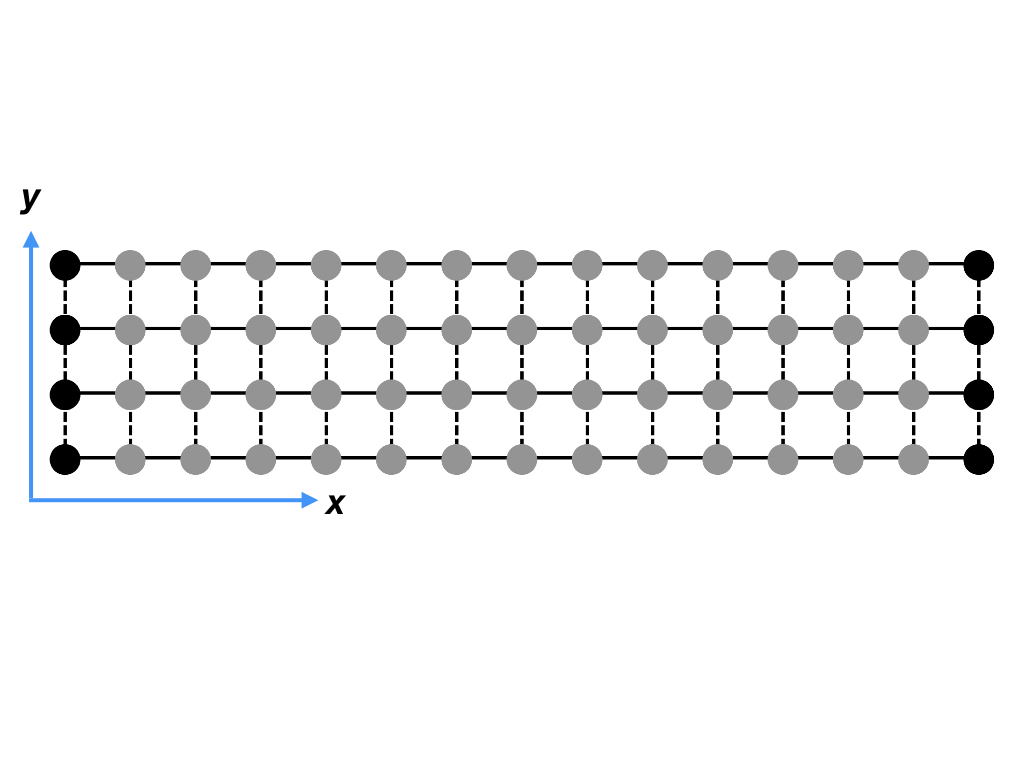}
\vspace{-.8in}
\caption{(Color online) A sketch of a finite-size strip of size $N_x \gg N_y$. The black sites denote the short edges of the system where the Majorana modes would be localized. This system can be thought of as a set of 1D wires coupled in the $y$-direction.}
	\label{fig6}
\end{figure}
In what follows we consider finite-size systems with open boundary conditions in both directions (see Fig.~\ref{fig6}), and we analyze their properties numerically by looking at the lowest energies eigenstates obtained using the MathQ code \cite{matq}. We take the magnetic field to be uniform and perpendicular to the plane of the lattice, $\vec{B}(\vec{r})=B \hat{z}$. For our analysis we ignore the orbital effects of the field. The interesting question is what happens when the transversal and longitudinal directions become comparable, in which situation we should expect that the system has effectively a single boundary and the Majorana states on the opposite edges may hybridize and destroy each other. 
%
\begin{figure}
\center
\includegraphics*[width=0.5\columnwidth]{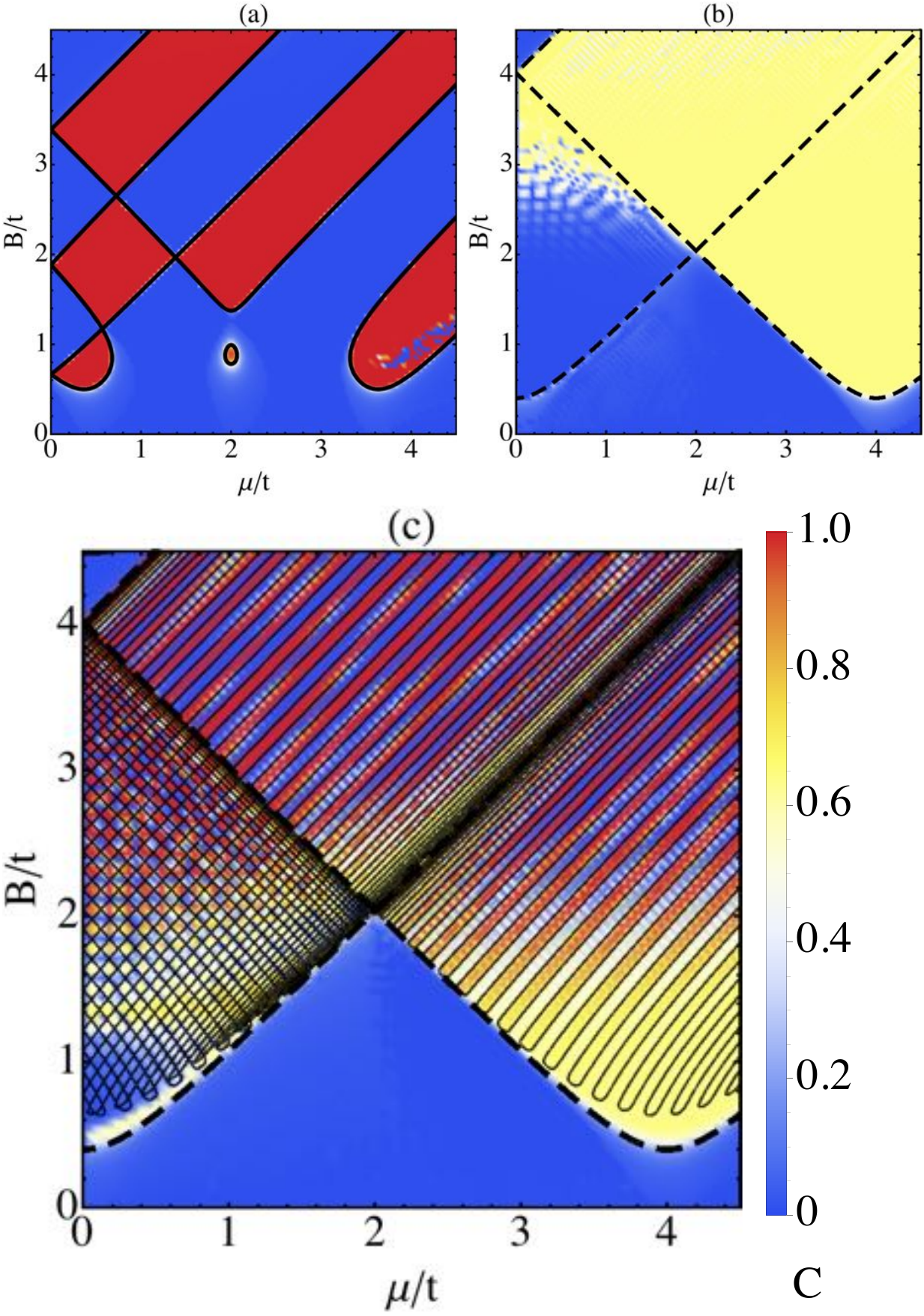}
\caption{(Color online) The phase diagram as a function of $B$ and $\mu$ for finite-size systems of (a) $3\times201$, (b) $101\times101$, and (c) $51\times201$. This has been calculated using  $C$, the renormalized sum of the MP vector of the lowest energy state over half of the system (see main text). In all the examples $\Delta=0.4t$ and $\alpha=0.5t$. The analytical topological phase boundaries for quasi 1D (solid lines) and 2D (dashed lines) are also plotted.}
\label{figure3}
\end{figure}


For a system in which the length is much larger than the width $N_y\ll N_x$ we expect to recover the checkerboard phase diagram calculated analytically in \cite{Sedlmayr2016}, and this is indeed the case, the topological regions predicted by the MP criterion (red in Fig.~\ref{figure3}(a)) correspond exactly to the topological regions predicted in \cite{Sedlmayr2016}. For a wider system  we can see that the phase diagram calculated analytically is recovered very well at large values of B, however, at small values of B, the value of $C$ is neither $0$, nor $1$, but it shows an intermediate value (regions depicted in yellow in the phase diagram). For fully square systems (Fig.~\ref{figure3}(b)) the entire topological phase is characterized now by $C\approx 0.7$. We should note that all the states corresponding to the regions with $C<1$ have non-zero but very small energies, thus making it impossible to interpret them as Majoranas or not based solely on the energy criterion. 

To understand the nature of these states we focus on the local structure of the MP. A Majorana state must have an integral of the MP  over a spatial region $\mathcal{R}$ equal to the sum of the LDOS over the same region. This can be achieved only if at each site the MP is equal to the DOS (perfect local Majorana character), as well as if the MP local vectors are all aligned inside $\mathcal{R}$ (`ferromagnetic' MP structure).

\begin{figure}
\center
\includegraphics*[width=0.5\columnwidth]{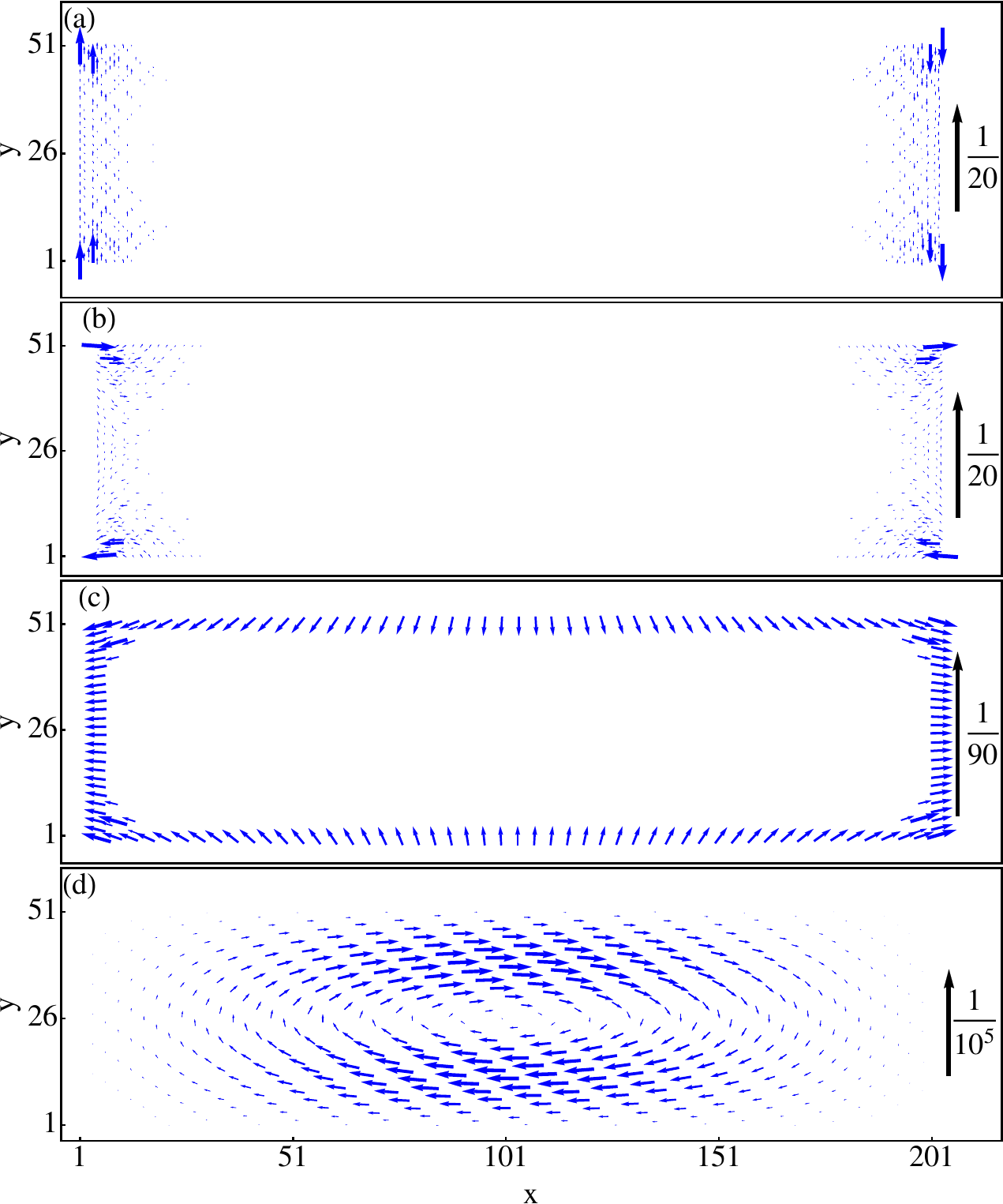}
\caption{(Color online) MP for the lowest energy states for an open system of $51\times201$ with $\Delta=0.3t$ and $\alpha=0.5t$.  We plot (a) an MBS $\mu=3.5t$ and $B=2.2t$ (corresponding to the red phase in Fig.~\ref{figure3}(c)), (b) a non-MBS edge state $\mu=3.5t$ and $B=2.3t$ (blue phase in Fig.~\ref{figure3}(c)), (c) a quasi-Majorana state $\mu=3.5t$ and $B=t$ (yellow phase in Fig.~\ref{figure3}(c)), and (d) a trivial bulk state $\mu=0.5t$ and $B=5t$ (blue phase in Fig.~\ref{figure3}(c)).}
\label{figure4}
\end{figure}

\begin{figure}
\center
\includegraphics*[width=0.7\columnwidth]{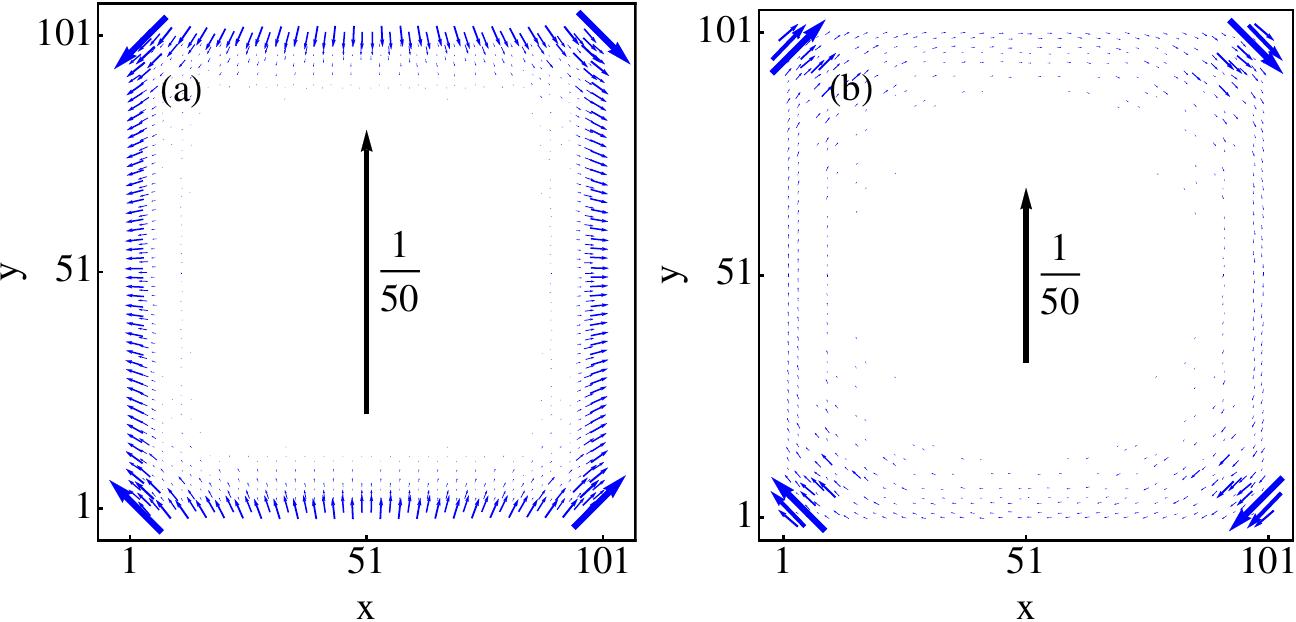}
\caption{(Color online)  MP for lowest energy state for open system of $101\times101$ with $\Delta=0.3t$ and $\alpha=0.5t$.  (a) $\mu=4t$ and $B=t$, and (b) $\mu=2t$ and $B=3t$, both in the 2D topologically non-trivial phase.}
\label{figure5}
\end{figure}

In Figs.~\ref{figure4}-\ref{figure5} we plot the MP for a variety of different low energy states. In Fig.~\ref{figure4} we plot the MP vector for a  $51\times201$ system with $\Delta=0.3t$ and $\alpha=0.5t$. Fig.~\ref{figure4}(a) corresponds to the red (topological) phase in Fig.~\ref{figure3}(c) ($C=1$), with $\mu=3.5t$ and $B=2.2t$. We should expect to have two MBS confined at the two narrow ends of the ribbon, and we note that this is indeed the case,  we can see the formation of two `ferromagnetic' states localized at the two ends of the wire, having opposite MP vectors. 

Fig.~\ref{figure4}(b) corresponds to the blue phase in Fig.~\ref{figure3}(c), with $C=0$. We take $\mu=3.5t$ and $B=2.3t$, and the corresponding state, while being localized on the edges, is indeed non-Majorana, since the integral of the MP over half of the wire is exactly zero.

In Fig.~\ref{figure4}(c) we focus on the more puzzling $C<1$ phase, denoted in yellow in Fig.~\ref{figure3}(c). We take $\mu=3.5t$ and $B=t$. We note that the corresponding lowest energy state is also an edge state, extending over the entire contour of the system. Moreover, the state is not `ferromagnetic', but exhibits a small uniform variation between adjacent sites, making the total integral over $R$ finite, but not equal to $1$. We denote this state as a quasi-Majorana state.
 
 Finally in Fig.~\ref{figure4}(d) we show a state with $C=0$ for $\mu=0.5t$ and $B=5t$, which corresponds to the blue region outside the region delimited by the 2D topological lines. This actually correspond to a fully trivial state localized no longer on the edge but in the bulk, and for which the value of the MP is close to zero even locally. 

It appears that the existence of the `yellow' phase is due to the shape of the system, i.e to the fact that the length and the width become comparable. If the length is increased the yellow phase is diminishing and the phase diagram converges towards the phase diagram calculated analytically in the previous section. This observation has been made also for the Kitaev model \cite{Sedlmayr2015b}. However, we want to stress that in real experimental systems the existence of the intermediate phase with $C<1$ is possible, and it would correspond to the formation of low-energy subgap states which have locally a full Majorana character, but that are non-Majorana, due to their global lack of symmetry. It would be interesting to explore the connection between these quasi-Majorana states and the 'chiral' Majorana states described in Refs.~\cite{Potter2010,Alicea2012}. Another interesting question would be if such quasi-Majorana states have any topological characteristics, such as non-Abelian statistics, or atypical braiding properties, and if such states can be useful for example for quantum computation in a similar manner as the Majorana states. 
Note that a state with a $C=1$ is a Majorana, and certainly has non-Abelian statistics. It is not at all clear that the statistics of the quasi-Majorana states would be Abelian. These are peculiar states in-between fermionic and Majorana-like, and there exist to this point no work to investigate their statistics. Moreover there seems to be a fundamental difference between such states and the properly fermionic states which have a uniform local zero MP. The quasi-Majoranas have their weight divided into a few different regions, roughly disconnected from each other, each with $C$ different from zero. Manipulating such a state in one part of the system is not the same as manipulating the fermionic state, and the effect of such manipulation is also a very interesting question. 

An extreme situation is depicted in Fig.~\ref{figure5}(a,b), in which we plot the MP for two quasi-Majorana states for a square system. Interestingly enough these two states have different MP characteristics, they are either localized roughly at the corners, or along the contour of the system. Note that for a fully square system the corner MP vectors are oriented at roughly $90$ degrees angle with respect with each other, giving rise to an integral of $C=\sqrt{2}/2\approx 0.7$, which corresponds well to the `yellow' in Fig.~\ref{figure3}(b).

\section{Conclusions}
We have introduced the Majorana polarization vector and we have shown that it can be used as a local order parameter to characterize a given eigenstate. We have shown that using a criterion based on the MP (the sum of the MP vectors over a given region is equal to the sum of the LDOS) allows one to distinguish Majorana states from non-Majorana states. We have implemented this criterion to one-dimensional systems and finite-size two-dimensional systems. We have proved that using such a criterion is crucial in studying the character of very-low-energy states as it allows one to distinguish Majorana states for example from quasi-Majorana states that are locally Majorana, but for which the MP vector is not aligned inside a given region.

\section*{References}

\bibliography{biblio_MF}

\end{document}